\documentclass[10pt,journal,compsoc]{IEEEtran}

\usepackage[english]{babel}
\usepackage[utf8]{inputenc}
\usepackage{caption}
\usepackage{makecell}
\usepackage{tabularx}
\usepackage{xcolor}
\usepackage{url}

\ifCLASSOPTIONcompsoc
  \usepackage[nocompress]{cite}
\else
  \usepackage{cite}
\fi

\ifCLASSINFOpdf
  \usepackage[pdftex]{graphicx}
  \graphicspath{../figures/}
  \DeclareGraphicsExtensions{.pdf,.jpeg,.png}
\else
  \usepackage[dvips]{graphicx}
  \graphicspath{{../eps/}}
  \DeclareGraphicsExtensions{.eps}
\fi
\begin{document}
%
\title{Paperclickers: Affordable Solution for \\ Classroom Response Systems}
%
%
%
%


\author{
        Eduardo Oliveira,
        Jomara~Bindá,
        Renato Lopes,
        Eduardo Valle*
        
\thanks{
J. Bindá was, and E. Oliveira, and E. Valle are with RECOD Lab --- University of Campinas (UNICAMP), Campinas, SP, Brazil.\protect\\
J. Bindá was, and E. Oliveira, R. Lopes, and E. Valle are with School of Electrical and Computer Engineering --- University of Campinas (UNICAMP), Campinas, SP, Brazil.\protect\\
J. Bindá is now with Center for Human-Computer Interaction --- College of Information Sciences and Technology --- The Pennsylvania State University, State College, PA, USA.\protect\\
\** Corresponding author: dovalle@dca.fee.unicamp.br, mail@eduardovalle.com.\protect\\
Manuscript received Month day, year; revised Month day, year.}}

\IEEEtitleabstractindextext{%
\begin{abstract}

We propose a low-cost classroom response system requiring a single mobile device for the teacher and cards with printed codes for the students. We aim at broadening the adoption of active learning techniques in developing countries, offering a tool for easy implementation. We embody the solution as a smartphone application, describing the development history, pitfalls, and lessons learned that might be helpful for other small academic teams. We also described the results of the first round usability tests we performed on the first prototype, and how the affected the current version of the software. A beta release version is currently available for the public at large.

\end{abstract}

\begin{IEEEkeywords}
Active Learning, Clickers, Audience Response System, Classroom Response System, Image Processing, App Development.
\end{IEEEkeywords}}

\maketitle

\IEEEdisplaynontitleabstractindextext

%
\IEEEpeerreviewmaketitle

\IEEEraisesectionheading{\section{Introduction}\label{sec:introduction}}
\IEEEPARstart{A} Classroom Response System (CRS) allows polling the students in real time, facilitating active pedagogical practices, which improve learning. They are often implemented as ‘clickers’, small infrared or radio-frequency remote-controls the students use to record their answers.


Deploying ‘clickers’ involve many costs: acquiring the devices, installing the receivers, training the personnel, and managing the operation (e.g., dealing with batteries, etc.). The total cost  is often unfeasible for schools in developing countries --- in Brazil, for example, they are virtually absent from public schools. 

To address that issue, we propose the CRS as a smartphone app, using the camera to scan the students' answers on printed codes. Our solution is easy to use, forgoes network connectivity, and requires a single hardware device per classroom (which may be the phone the teacher already owns).

The idea of using cheap printed codes to get student feedback is not novel~\cite{vickrey2015PI}~\cite{cross2012low}~\cite{miura2012device}, but as always for successful implementation, the devil is in the details. As far as we known, paperclickers is the only existing solution in the intersection of being an academic work, having its entire source code publicly releases, and being released for download on user's devices for actual, practical use. Being on that intersection allows future contributors to easily test hypotheses and add improvements to paperclickers, with real-world impact to users.

In this article we discuss (1) the architecture of our solution; (2) the technical experiments that lead us to choose TopCodes~\cite{horn2012topcode} to use on the students' cards, and the several adaptations we had to introduce to recognize dozens of codes in the uncontrolled environment of a classroom; (3) the user experiments on the first prototype; (4) the improvements and adaptations leading to the released app. We also summarize the development history, with the lessons we learned, which may be helpful for other academic teams venturing into app creation. 


The remaining text is organized as follows: in section 2, we establish the pedagogical context, with the expected benefits of CRSs; in section 3, we review available CRS technologies; in section 4, we present the development processes for our solution, from initial prototype until beta release, passing through technical and user-interface experiments. In section 5, we conclude with a short discussion about our current work on the pedagogical components.

\section{Pedagogical Context} \label{sec:relatedwork}
Feedback is a powerful tool --- if not \emph{the most} powerful tool --- for effective teaching and learning, especially when it flows “from the student to the teacher”, that is, “when teachers seek, or at least are open to, feedback from students as to what students know, what they understand, where they make errors, when they have misconceptions, when they are not engaged”~\cite{hattie2009visiblelearning}. However, to deliver such power, feedback has to provide timely information about the process of learning. Instead of focusing on assignments' correctness, or on students' performance --- as praising, punishing or giving external rewards do --- feedback needs to fill “the gap between what is understood and what is aimed to be understood”. 

Peer Instruction (PI)~\cite{mazur1997peer} is a teaching method that institutes regular feedback to probe student understanding. It differs from traditional lectures, introducing topic-specific questions (“concept tests”) designed to reveal conceptual misunderstandings. The procedure:

\begin{itemize}
    \item Start class by presenting a topic;
    \item Ask a conceptual question (“concept test”) to check students’ understanding;
    \item Wait a moment for the students commit to an answer, then make a poll;
    \item If a large majority of the answers is right, clarify the few wrong answers and move on. If a large majority of the answers is wrong, present the topic again;
    \item If the answers are  divided, withhold the right answer, and have the students discuss the topic in small groups. After a set time, poll again. Almost invariably, the class will converge towards the right answer;
    \item Repeat for the next topic.
\end{itemize}

Due to its flexibility --- it adapts to any topic or teaching style --- PI has been extensively adopted.
The concept tests can be collected, organized, reused, and shared, within and across institutions~\cite{crouch2001peer, mazur1997peer}.

Vickrey et al.~\cite{vickrey2015PI} review 56 studies at STEM colleges, which report PI in large (\textgreater50 students) classrooms. They find PI offers measurable learning gains, twice larger than those offered by traditional lectures; PI also improves problem-solving, some of the studies reporting increased “ability to answer questions designed to measure mastery of material” and “to solve novel problems (i.e. transfer knowledge)”, as well as improved “quantitative problem-solving skills”. Finally, PI reduces attrition rates, with reduced dropout, and less failures.

PI's conceptual tests require polling the students, which, in a pinch, can be done by show of hands. But, as any teacher can report, students are tempted to follow the majority, or simply refuse to raise their hands for any of the presented alternatives. Giving the students cards of different colors, and asking them to raise one of them (corresponding to a yes/no or multiple-choice answer) is a better alternative: still, some students will attempt to follow the majority. Those “low-tech” alternatives only provide the teacher a rough aggregate feedback from the classroom, they preclude later analysis of individual answers.



\subsection{Impacts of Classroom Response Systems}

Several studies~\cite{yourstone2008classroom, caldwell2007clickers} suggest that CRSs make the students more attentive: knowing that they will be polled make them better prepared for classes. 
Other studies~\cite{kay2009examining, caldwell2007clickers} mention frequent and positive interactions  when using CRSs during discussion sessions, reporting better articulated thinking~\cite{beatty2005transforming}, since students have to commit to an answer, and, if required, defend it (when CRS are used in combination with PI).

There are also potential benefits of CRSs for instructors, since they can analyze learning in real time~\cite{yourstone2008classroom} and adapt the lecture accordingly~\cite{caldwell2007clickers, cutts2006practical}. Some~\cite{yourstone2008classroom, caldwell2007clickers} suggest that such immediate feedback increases learning performance.

Hunsu et al.'s meta-analysis~\cite{hunsu2016ARSeffect} investigated the cognitive and non-cognitive effects of CRSs, compared to conventional lectures, comprising over 50 selected articles, with a total of 26,085 participants. They considered 111 independent learning outcomes, coded from all the variables present in the original studies, split in 86 cognitive and 25 non-cognitive outcomes. The meta-analysis shows that CRSs improve students' participation in large lecture halls (a common, and particularly challenging, environment for STEM courses). It also reveals small, but positive, cognitive outcomes of those technologies, specially on higher-level learning goals, like knowledge transfer, or knowledge application.

Despite those positive results, CRS implementation is challenging. Technological problems may occur (devices not working properly, communication problems)~\cite{siau2006use}. Applying CRSs without pedagogical preparation may not benefit the learning process, and can even harm the positive perception about those devices within classes. Preparing proper questions --- those which identify misconceptions --- is hard; if a standard set of questions is not available for using the CRS, the instructor will have the burden of choosing them for each lecture~\cite{caldwell2007clickers}.

Furthermore, CRS benefits only appear if classroom culture supports student engagement and interaction. Students must feel their answers are important, in order to engage, and to take the procedure seriously. Otherwise, adverse situations may arise, like giving wrong answers on purpose~\cite{siau2006use}.

\section{Classroom Response Systems} \label{sec:crs}

CRSs started with Classtalk~\cite{beatty2005transforming}, which used graphing calculators to allow students communicate. In addition to dedicated ‘clicker’ devices, current solutions include “bring your own device” (BYOD), and solutions based in image-processing. Table~\ref{tab:1} presents and compares those solutions.

Vickrey et at. literature review~\cite{vickrey2015PI} concludes PI can be effectively implemented with clickers or any other tools --- suggesting image-processing clickers solutions are as effective as any other polling methods.


\begin{table*}[t]
    \begin{center}
        \begin{tabular}{l l l}
            \hline
             Method &  Advantages &  Disadvantages\\
            \hline
            \makecell[l]{“Low-tech” alternatives\\(show of hands, color cards)}&
            \makecell[l]{
                Very low-cost\\
                Available immediately everywhere\\
                Very easy to use
            }&
            \makecell[l]{
                Classmates tend to “follow the majority”\\
                Individual answers unrecoverable\\
                Only multiple-choice answers possible
            }\\

            \hline
            \makecell[l]{Dedicated hardware\\(‘clickers’)}&
            \makecell[l]{
                Wide commercial availability\\
                Classmates cannot see answers\\
                Instructor recovers individual answers\\
                Moderate to complex answers possible
            }&
            \makecell[l]{
                High direct and indirect costs\\
                Complex training required for teachers
            }\\

            \hline
            \makecell[l]{Software on students' devices\\(BYOD)}&
            \makecell[l]{
                Good commercial availability\\
                Classmates cannot see answers\\
                Instructor recovers individual answers\\
                Very complex answers (e.g. drawings) possible\\
                Low-cost for institutions
            }&
            \makecell[l]{
                High-cost for students\\
                Devices can be distracting\\
                Requires reliable network infrastructure\\
                Training required for teachers and students
            }\\

            \hline
            \makecell[l]{Software on teachers' device +\\cards with barcodes for students\\(image processing)}&
            \makecell[l]{
                Low-cost for students and institution\\
                Classmates cannot see answers\\
                Instructor recovers individual answers\\
                Simple training required for teachers,\\virtually no training for students
            }&
            \makecell[l]{
                Few (mostly experimental) solutions\\
                Only multiple-choice answers possible\\
                Requires line of sight to each student
            }\\

            \hline
        \end{tabular}
    \end{center}
    \caption{Summary of classroom response system technologies. Image-processing CRSs --- like paperclickers --- are the only ones at the intersection of low cost, simplicity, anonymity to classmates, and trackability of answers by instructors. }
    \label{tab:1}
\end{table*}

\subsection{Clickers}
In their most usual form, CRSs use ‘clickers’, one small infra-red or radio-frequency transmitter per student, and one receiver per classroom\cite{caldwell2007clickers}. When the instructor poses a question, the students use the transmitters to answer it, and the receiver instantly  records and tabulates those answers. The instructor can display the results for immediate action, save the data for later analysis, or both~\cite{beatty2005transforming}.

‘Clicker’ solutions count with wide commercial availability and professional support. If correctly implemented and well-maintained, they allow CRSs to run smoothly and non-intrusively~\cite{caldwell2007clickers}. The commonest devices allow only multiple-choice or yes/no answers, but recent (and more expensive) versions allow numbers, words, or short phrases.

Those solutions, however, are expensive. Ideally, each student should own their transmitter (to avoid the time-consuming hassle of distributing and collecting the devices at each class), and receivers must be available for each classroom or, if they are portable, teacher. Teachers have to be trained to use the transmitters~\cite{beatty2005transforming}, and support must be readily available to solve problems~\cite{caldwell2007clickers}, minimizing class disruption. The system needs continual maintenance to work optimally; in particular, it is critical to establish who will be responsible for the devices' batteries (students, teachers, or technicians), and ensure that  responsibility is taken seriously.

Due to those costs and inconveniences, CRSs based on ‘clickers’ are very challenging to implement in developing countries.

\subsection{Bring Your Own Device}

“Bring Your Own Device” (BYOD) --- a model in which students use their own smartphones, tablets, or even laptop computers --- replace hardware ‘clickers’ by software applications that transmit the answers through WiFi or mobile data~\cite{stavert2013byod}. There is a good offer of commercial BYOD software, sometimes from the same vendors of hardware ‘clickers’, even allowing hybrid solutions where both are supported at once.


CRS software usually has modest hardware requirements, allowing to apply BYOD to a wide range of devices, including obsolete/donated hardware. BYOD also allows very flexible forms of polling, that no other solution provides, including mini-essays, points and graphs in Cartesian coordinates, or even freehand drawings.

Teachers and students still need to be trained to use the system, and there still needs to be technical support, especially to solve connectivity issues. Indeed, network quickly becomes a bottleneck --- both when using public (mobile data) or local (WiFi) infrastructure --- as it is challenging to support many simultaneous connections. Ensuring the usage of the devices remains on purpose and productive is an additional challenge, as students may turn to leisure texting or web browsing.

BYOD assumes there is good network infrastructure, and that each student owns at least a smartphone, and that they feel safe to bring it to school~\cite{stavert2013byod}.  Those assumptions are far from obvious in developing countries, especially in schools serving disfavored communities. For example, the Brazilian Internet Steering Committee shows~\cite{CGIbr2014ITCeducation} that as recently as 2014, even urban schools in Brazil had connectivity issues: although 93\% of those schools had some access to the Internet, only 41\% of them granted access to students. In addition, connectivity was often low-speed and unreliable.

\subsection{Image Processing}

Image-processing solutions minimize costs by giving the students passive devices, usually cards with especial colors or codes, and keeping all active processing into a single device, which remains with the teacher. Most often, the students rotate use a card printed with a special 2D barcode, which serves both as a location and orientation marker, and as a unique ID for each student. The students can answer multiple-choice questions by rotating the cards.

Amy and Amy patented a low-cost optical polling framework~\cite{nolan:2011:biblatex} with a generic computing element that recognizes the orientation of fiducial marks on printed cards. The proposed solution is available as a smartphone app --- Plickers\footnote{https://plickers.com/}, which currently accommodates up to 63 students, which must enroll on a web-based system. Fiducial markers on printed cards had already appeared on previous works, e.g., the augmented reality system ARTag~\cite{fiala2005artag}. Amy and Amy innovate by exploiting them for low-cost CRSs. 

Cross et al.~\cite{cross2012low} also proposed a system that recognized the orientation of printed cards with unique IDs for each student, which they called  ‘qCards’. The teacher captures the answers with an off-the-shelf webcam mounted on a laptop, with software to recognize, tabulate, and display the results. The authors ran initial trials on secondary schools in Bangalore, India, with 99.8\% recognition accuracy, and 97\% captured responses in a 25-student classroom.

Miura and Nakada presented similar work~\cite{miura2012device}: they used printed cards with fiducial markers as codes, and a similar setup of camera, and PC containing the software. Their system recognizes three rotational parameters for each cards (roll, pitch and yaw), allowing students to select one of many possible multiple-choice answers in a screen. A preliminary experiment, with 19 students succeeded in tracking 18 markers.

The solution proposed by Gain~\cite{gain2013using} approaches CRSs by colored blocks printed on cardboards and a camera-phone to capture images. Students select answers by picking different colors. They report 85\% recognition accuracy in a medium-size class (up to 125 students). Although the system forgoes peer anonymity, and the possibility to track responses to individual students for later analysis, it is the only image-processing system tested in classes that big.

Finally, Ito and Miura~\cite{ito2015portable} experimented a portable version of that previous system, recognizing the same fiducial marks in an tablet computer, including the capability of detecting the response printed card bending amount as an additional input mechanism, which could encode the student mood.

As far as we known, paperclickers is the only existing image processing CRS solution at the intersection of being an academic work, having its entire source code publicly released, and being available for download on user's devices for actual, practical use. Being on that intersection allows future contributors to easily test hypotheses and add improvements to paperclickers, with real-world impact to users.

\section{Development of Paperclickers} \label{sec:proposal}

Paperclickers aims at lowering the costs of CRSs, and thus fostering their adoption in Brazil and other developing countries. We follow the image-processing model, using fiducial markers on the students cards, and keeping the software in a mobile device (smartphone or tablet) that stays with the teacher. The students answer multiple-choice questions by rotating the cards into one of four orientations.

Paperclickers has very modest requirements: a single mobile device per classroom (which can be the teacher's personal smartphone), and no Internet connection. Cards must be printed and distributed to the students, but the cost is low, a few cents per card. The solution is very easy to setup and use, reducing costs of installation and training.

The main limitation, as we will see, is the number of students. There is a hard limit of 99 codes, but since there must be an unencumbered line-of-sight between the codes and the device, the practical limit may be lower. A lesser limitation is that only multiple-choice questions are possible, but that limitation is not a problem for active-learning techniques like PI.



\subsection{Initial Development}

Paperclickers' first design was similar to Cross et al.'s~\cite{cross2012low}, but exchanging the PC+webcam setup by a mobile device with embedded camera. In both designs, the instructor uses the camera to film the class, as the students hold up cards with their answers --- using four different orientations to pick among four possible answer choices. The system tabulates the answers in detail, or summarizes them in a graph.

\begin{figure}[ht]
    \centering
    \begin{minipage}{.5\textwidth}
        \centering
        \setlength{\fboxsep}{0pt}\fbox{\includegraphics[width=.9\textwidth]{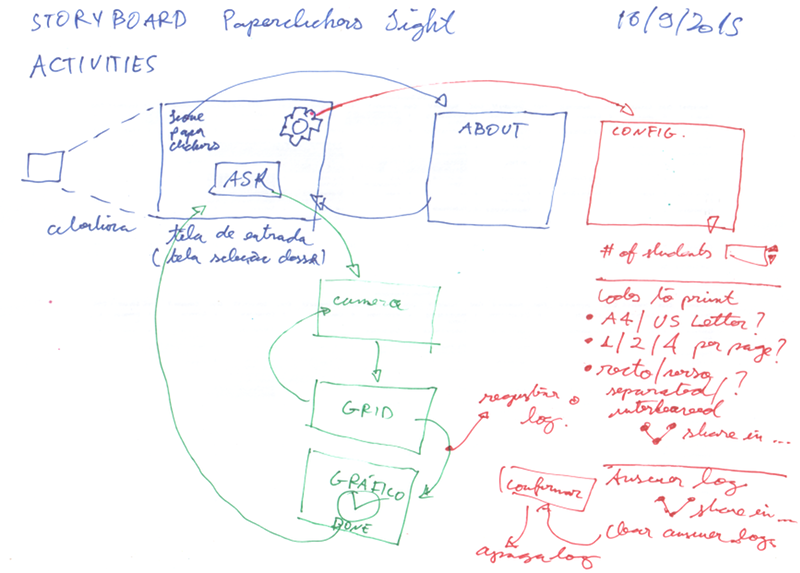}}
        \captionof{figure}{Actual storyboard from the second development cycle. Practical, informal, and easy to implement, storyboards proved effective for planning and communicating all design decisions among team members.}
        \label{fig:storyboard_2nd_cycle}
    \end{minipage}
\end{figure}

Development started with brainstorming sessions to storyboard the application use cases, its interface, and its behavior. The storyboards were the main planning tool for the development --- they provided a good compromise between our desire for an informal, lean process, and the need to design the application, and document and communicate the decisions among team members. The implementation was guided by the decisions made during storyboarding (figure~\ref{fig:storyboard_2nd_cycle}).

The storyboards described the following workflow:

\begin{enumerate}
    \item \textit{Opening the class}: the instructor selects the class identifier to open a session;
    \item \textit{Starting a question}: to start polling the students, the teacher select the question number, for later identification (those numbers autoincrement, but can also be changed manually);
    \item \textit{Image capture}: the camera activates and the instructor scans the entire class, capturing the answers;
    \item \textit{Results screen}: the application shows each student answer; if there remain unidentified answers, the instructor can go back to the image capture stage, or input the answers manually;
    \item \textit{Chart screen}: the application summarizes the result in a pie chart.
\end{enumerate}

Initial design envisioned an offline module to manage classroom definitions: class IDs, and their enrolled students. During prototyping we managed that information manually, directly on the application internal data files.

During initial tests, we added two new features: a separate flow for roll-call  --- letting the teacher to do an initial roll-coll in the classroom to later better identify absent students from unrecognized codes; and an augmented reality feedback on the answers capture screen, to let the user identifying recognized cards (Figure~\ref{fig:old_answersScan}).

\subsection{Recognizing Students' Answers}

At first, for the students cards, we chose QR Codes (ISO/IEC 18004)~\cite{ISO18004}, which can store many characters, have error correction, and have location patterns that easily establish their orienation~\cite{belussi2013fast}. We developed a prototype mobile app in Android\footnote{http://www.android.com/}, using the ZXing\footnote{ https://github.com/zxing/} open source library to generate and decode the QR Codes.

QR Codes would allow unique codes for each student across an institution, or even across an entire education system, since it can easily store dozens of digits. However, we found next to impossible to adapt them to our use case. QR Codes are optimized for recognition and decoding of a single code at close range, while we needed to find and decode dozens of codes across different distances from the camera, in a single photo. Although we obtained moderate success after several modifications in ZXing code, the decoding was still too slow to be used on a video stream, and had to employ still pictures. Worse yet: recognition accuracy --- even on ideal situations of illumination and occlusion --- was unacceptable.

Giving up on storage capability, a natural solution presented itself from the literature of motion capture and augmented/virtual reality: fiducial marks, which are engineered to be recognized in real-time in a video stream. We selected TopCodes\cite{horn2012topcode}, originally designed to create virtual objects for the Tern tangible programming environment~\cite{horn2007designing}. We still had to modify the recognition/decoding code for our purposes (Section~\ref{sec:second_dev_cyvle}), but after such fine-tuning they proved extremely robust and very fast. On the other hand, TopCodes support only 99 unique identifiers, making possible to give unique ids for each student only in small to moderate-large classes. Using on large institutions is not impossible, but requires the additional burden of keeping tables to translate class IDs into institution IDs for each class.

\subsection{User Experiments}


We applied usability testing on a small subject sample (N = 11), employing direct observation, user and interaction recording, semi-structured interviews and a questionnaire. Although we have found relevant usability issues, we concluded that the proposed prototype was adequate as CRS. Usability tests are effective when compared to the field testing~\cite{kaikkonen2005usabilitytests}, and can be cost effective with a small number of subjects: Nielsen Normal Group\footnote{How Many Test Users in a Usability Study - https://www.nngroup.com/articles/how-many-test-users/} points out 5 samples can provide an important usability insight. We detail the user-interaction/interview findings in Table~\ref{tab:user_device_interaction_findings}, and summarize the questionnaire findings in Table~~\ref{tab:user_questionnaire_findings_compilation}.




We tested two use cases, covering typical usage:

\textit{First use case --- starting the class: launching the app and performing a roll call.} The teacher starts the paperclickers application on her mobile device, and then chooses the class to start. After the students arrive, she touches the roll call icon. She asks the students to raise their cards (in any orientation) and films the entire class. She touches the screen to advance to the result screen. The application shows the roll call result, if some student is present but has not been detected, the teacher can correct the status by tapping the corresponding icon on the device screen. The teacher touches the "done" icon to finish the roll call procedure.

\textit{Second use case --- asking a question: during class the teacher polls the students.} The teacher wants to verify the students' understanding of a given topic. He opens the paperclickers application and touches the "Start" button. He chooses a number for the question. He enunciates a multiple-choice question, and waits for the students to commit to an answer; then he asks the students to lift their cardboards to indicate the chosen answer, using the letters printed on the back of the card as cue (the chosen letter must be upright). The teacher touches the "Collect answer" button to advance to the scan screen. After capturing the answers, he touches the screen to advance to the detailed results screen. If many answers are missing, he can choose to go back to the scan screen, and then the application would merge the results from both scans. He can also touch the icons to manually change the answers. If the answers are correct, the teacher can touch the graph icon to advance to the summary results screen, that shows a pie chart. At this point, the teacher can touch the "Try again" button to return to the scan screen or the "New question" button to pick the number of the next question.

The prompts we gave to the test subjects were much less directive than the detailed scripts shown above. For example, we would ask the users "please, start a roll-call" not "touch the roll-call icon", and would observe if the users were able to figure the interface by themselves. Only when the users were blocked for several minutes we would give a direct prompt.

Our experiments ended up reflecting the DECIDE framework, as described by Preece et al.~\cite{preece2002interaction}, which establishes a checklist comprised of 6 steps to guide the evaluation preparation: \textit{“Determine the overall goals that the evaluation addresses; Explore the specific questions to be answered; Choose the evaluation paradigm and techniques to answer the questions; Identify the practical issues that must be addressed, such as selecting participants; Decide how to deal with the ethical issues; Evaluate, interpret, and present the data”}.

\emph{Determine the goals:}
\begin{itemize}
    \item The solution offers a convenient and consistent user interaction, satisfying the user;
    \item The overall design allows the user to perform easily and efficiently the required tasks, suitable to the users' real problem;
    \item It is easy to learn how to use the solution.
\end{itemize}

\emph{Explore the questions:}
\begin{itemize}
    \item Is the \textit{answers capturing} screen, with the augmented reality codes visualization, legible and readily explainable to the user?

    \item Are the \textit{detailed answers visualization} and \textit{answers chart} screens legible and readily explainable to the user?

    \item Is the screen navigation throughout the application consistent and does it match user expectations?
\end{itemize}

\emph{Choose evaluation paradigm and techniques:}


\begin{itemize}
    \item Direct observation and monitoring;

    \item Recording of the user with a video camera;

    \item Recording of the user interface with screencap;

    \item Semi-structured interview after the experiments, asking about the overall experience and main difficulties;

    \item Questionnaire after the experiments, exploring quantitative information on the user satisfaction and understanding about specific visual elements.
\end{itemize}



\emph{Identify practical issues:}

\begin{itemize}
    \item The tests would be applied on volunteer participants older than 18-year old. We did not put any other restrictions, but since we announced the test in the University, we knew we would be mainly sampling graduate and undergraduate students. Although a broader audience would be desirable, we considered that a sample of students was aligned with at least part of the target users, providing thus already valuable feedback. Broader audiences could be addressed later, on a refined product.

    \item We defined test scripts, to be used as instructions during the experiment, for both the applier and the test subjects.

    \item The experiment procedure considered only one participant using the application; we fixed the cardboards on the backrests of the class' seats, simulating the students answering performance during the class. A research team member would follow for support.

    \item We defined a controlled environment, using two rooms at the University: the first for subjects using the app, and the second room for the semi-structured interview and filling the questionnaire.
\end{itemize}

\emph{Decide how to deal with the ethical issues:}

Although the proposed tests corresponded to the simple indoor usage of a smartphone application, without repetitive actions or sensitive data collection, we identified and planned to mitigate any ethical concerns.

\begin{itemize}
    \item The privacy of user information gathered throughout the usability tests, the interview and questionnaire, also including the recorded personal image; we created an Informed Consent Form to clarify our privacy agreement.

    \item Any discomfort the users might few performing the tests, related to shyness of begin recorded or interviewed, for example; we provided clear and formal recommendations about the absolute freedom the participants had to leave the experiment during any phase.
\end{itemize}

The experiment was approved by the Brazilian research ethics committee.




\begin{table*}[!t]
    \begin{center}
        \renewcommand{\arraystretch}{1.5}
        \begin{tabular}{ p{0.04\textwidth} | >{\raggedright\arraybackslash}p{0.15\textwidth} | >{\raggedright\arraybackslash}p{0.15\textwidth} | >{\raggedright\arraybackslash}p{0.15\textwidth} | >{\raggedright\arraybackslash}p{0.15\textwidth} | >{\raggedright\arraybackslash}p{0.20\textwidth}}
            \hline
            \textbf{Tester} & \textbf{Backward navigation} & \textbf{Roll call (task 1)} & \textbf{Answers scan (task 2)} & \textbf{Test script} & \textbf{Additional comments} \\
            \hline
            \centering 1 & Misleading “Camera close” message; Inconsistent behavior & Could not change student status & Didn't use “New question” for next question & & Might reduce the teachers/students relation \\
            \hline
            \centering 2 & Inconsistent behavior & Unable to execute roll-call feature & Didn't use “New question” for next question & \\
            \hline
            \centering 3 & & Could not start feature without help; Could not change student status & Looked for additional information on chart screen & Didn't understand how to answer script question & \\
            \hline
            \centering 4 & & Unclear presence/absence icons & Looked for additional information on chart screen & & Liked roll call feature agility \\
            \hline
            \centering 5 & & Problems to start roll-call feature; Detection problems & Detection problems; Looked for additional information on chart screen & Didn't understand how to answer script question & Problems to dismiss about screen \\
            \hline
            \centering 6 & Misleading “Camera close” message & Unclear roll-call icon; Detection problems & Detection problems; Didn't use “New question” for next question & & Application low speed; Missing back option in detection screens; Focused on specific usage scenarios \\
            \hline
            \centering 7 & Misleading “Camera close” message; Inconsistent behavior & Problems to understand roll call feature & Detection problems & & The single device requirement might not be low cost; Students having to keep big cardboard signs might be a problem \\
            \hline
            \centering 8 & & Problems to start roll call feature & Detection problems & & Scanning large classroom could be cumbersome \\
            \hline
            \centering 9 & Inconsistent behavior & Unable to execute roll call feature & Used "New question" but couldn't realize the question number auto increment & \\
            \hline
            \centering 10 & & Problems to start roll call feature & Detection problems & Didn't understood how to answer script question & Forced landscape orientation; Found inconsistent the ability to change student presence while changing answer; Asked for more than 4 answers' choices; Question about detection in real classrooms \\
            \hline
            \centering 11 & & Detection problems & Detection problems & & Problems to dismiss class selection list; Would be nice to have the question text; Its usage might distrait the students \\
            \hline
        \end{tabular}
    \end{center}
    \caption{Recording the interaction of user with the app provided the most actionable information on the usability tests test, which were positive regarding application usage, but revealed that some features and navigation were confusing to users.}
    \label{tab:user_device_interaction_findings}
\end{table*}

\begin{table*}[!t]
    \begin{minipage}[t]{0.5\linewidth}
        \renewcommand{\arraystretch}{1.5}
        \begin{tabular}[t]{ >{\raggedright\arraybackslash}p{0.50\textwidth} | >{\centering\arraybackslash}p{0.05\textwidth} | >{\centering\arraybackslash}p{0.05\textwidth} | >{\centering\arraybackslash}p{0.05\textwidth} | >{\centering\arraybackslash}p{0.05\textwidth}}
            \hline
            \textbf{Element} & \textbf{L/A} & \textbf{In} & \textbf{D} & \textbf{DU} \\
            \hline
            Application forced landscape & 2 & 3 & 6 & \\
            \hline
            Initial screen -- class selection option & 8 & & & 3 \\
            \hline
            2\textsuperscript{nd} screen -- question selection option & 8 & & 1 & 2 \\
            \hline
            2\textsuperscript{nd} screen -- question auto increment & 4 & 7 & & \\
            \hline
            2\textsuperscript{nd} screen -- roll call separated from answers scanning & 6 & 2 & 3 & \\
            \hline
            2\textsuperscript{nd} screen -- roll call icon & 5 & & 3 & 3 \\
            \hline
            Scanning screen -- understood augmented reality cardboard indications & 8 & 3 & & \\
            \hline
            Scanning screen -- augmented reality cardboard indications & 6 & 5 & & \\
            \hline
            Scanning screen -- found augmented reality feedback slow & 6 & 5 & & \\
            \hline
            Scanning screen -- cardboards capture finalization method & & 3 & 8 & \\
            \hline
		    Roll call results screen -- easily understood & 9 & 2 & & \\
            \hline
		    Roll call results screen -- presence/absence icons understanding & 5 & & & 1 \\
            \hline
		    Roll call results screen -- would like to have student name or picture along presence/absence icons & 5 & 6 & & \\
            \hline
		    Roll call results screen -- easily understood presence/absence icons were clickable & 5 & 6 & & \\
            \hline
		    Roll call results screen -- screen closing icon & 5 & & 6 & \\
            \hline
            Detailed answers screen -- layout & 10 & 1& & \\
            \hline
            Detailed answers screen -- easily understood answers were clickable and could be changed & 6 & 1 & & 4  \\
            \hline
            Detailed answers screen -- understood “X” answer indication & 7 & 3 & & 1 \\
            \hline
            Detailed answers screen -- understood chart screen icon & 11 & & & \\
            \hline
            Detailed answers screen -- “back” icon & 8 & & 3 & \\
            \hline
        \end{tabular}%
    \end{minipage}%
    \begin{minipage}[t]{0.5\linewidth}
        \renewcommand{\arraystretch}{1.5}
        \begin{tabular}[t]{ >{\raggedright\arraybackslash}p{0.50\textwidth} | >{\centering\arraybackslash}p{0.05\textwidth} | >{\centering\arraybackslash}p{0.05\textwidth} | >{\centering\arraybackslash}p{0.05\textwidth} | >{\centering\arraybackslash}p{0.05\textwidth}}
            \hline
            \textbf{Element} & \textbf{L/A} & \textbf{In} & \textbf{D} & \textbf{DU} \\
            \hline
            Detailed answers screen -- correctly understood “back” icon would return to the scanning screen & & 2 & & 9 \\
            \hline
            Detailed answers screen -- understood “back” icon would return to the process beginning & 4 & 7 & & \\
            \hline
            Chart screen & 10 & & 1 & \\
            \hline
            Chart screen -- correctly understood “Try again” button would return to scan screen keeping the question & 7 & 1 & & 3 \\
            \hline
            Chart screen -- understood “Try again” button would return to detailed answers screen keeping the question & 1 & 1 & & 9 \\
            \hline
            Chart screen -- understood “Try again” button would return to the initial screen for class selection screen & 2 & 1 & & 8 \\
            \hline
            Chart screen -- correctly understood “New question” button would finalize the question and return to the question selection screen & 8 & & & 3 \\
            \hline
            Chart screen -- understood “New question” button would return to the answers scanning screen & 2 & & & 9 \\
            \hline
            Chart screen -- understood “New question” button would return to the detailed answers screen & 1 & & & 10 \\
            \hline
            Chart screen -- correctly understood “BACK” button would finalize the question and return to the initial screen for class selection & 3 & & & 8 \\
            \hline
            Chart screen -- understood “back” button would finalize the question and return to the question selection screen & 4 & & & 7 \\
            \hline
            Chart screen -- understood “back” button would finalize the question and return to the answers scanning screen & 1 & & & 10 \\
            \hline
            Chart screen -- understood “back” button would finalize the question and return to the detailed answers screen & 3 & & & 8 \\
		    \hline
        \end{tabular}%
    \end{minipage}%
    \caption{User questionnaire findings compilation: L = Liked; A = Agree; D = Disliked; DU = Didn't understand; In = Indifferent. Semi-structured interviews provided quantitative data, mostly reinforcing the usability test findings of the device interaction recording. We also captured users' preferences on other aspects, e.g., discontentment with the forced landscape orientation.}
    \label{tab:user_questionnaire_findings_compilation}
\end{table*}

\emph{Evaluate, interpret and present the data:}

\begin{itemize}
    \item The overall feedback was positive regarding the application usage and suitability to the proposed tasks: its main features were easily recognized and all the intended tasks successfully and easily performed by the majority of the testers;

    \item All the testers were able to successfully scan the class for results --- understanding the overall usability, including the augmented reality feedback --- as well as to read and manipulate the detection results in the detailed answers screen;

    \item We found two major usability issues about application convenience (roll call feature identification and initialization) and consistency (backward navigation throughout the application's screens);

    \item The user interaction recording provided the most valuable data, identifying the users' actions along with users' impressions, through touches or audible comments (also recorded); the post-test semi-structured interview provided important data to confirm the findings, adding overall impressions on the application proposal.
\end{itemize}

The recordings made clear the second screen was confusing and inconvenient: users were requested to perform a roll call --- reachable by an icon on the upper left corner --- but the main graphic elements were the "question number definition" box and its controls, and the "Collect answer" button (refer to figure~\ref{fig:old_questionSelection}). We can infer this confusion through the users reactions: some of them navigated backwards --- either through the device operational system "back" key or using the "home" icon --- and some of them even expressed audible surprise, giving the impression they were expecting something else.



We found consistency issues related to backwards navigation. Depending on the screen, backward navigation lead the user one, two or even three screens back. That was not only internally inconsistent, but also mismatched Android's expected behavior. One egregious example was the navigation from the chart screen, which represents the end of the scanning process: as visible in figure~\ref{fig:old_chart}, there were 4 different UI elements to navigate backwards, each one with a different behavior. Although those options gave users flexibility, they were very confusing.




During the semi-structured interview all the testers gave positive feedback on the application and usage scenario, but expressed a few concerns, like whether its usage might distract the students, might reduce intimacy between teachers and students distance, or might be unreliable on large classrooms.


The questionnaire data analysis (Table~~\ref{tab:user_questionnaire_findings_compilation}) was important mainly to confirm the same issues detected through the user interaction and interview, providing quantitative data regarding users' preferences on other aspects, like the fact most of the users disliked the forced landscape orientation of the app.




\begin{table*}[p]
    \begin{tabular}[t]{m{0.5\textwidth}m{0.5\textwidth}}
        \begin{minipage}{0.5\textwidth}
            \centering
            \setlength{\fboxsep}{0pt}\fbox{\includegraphics[width=.9\textwidth]{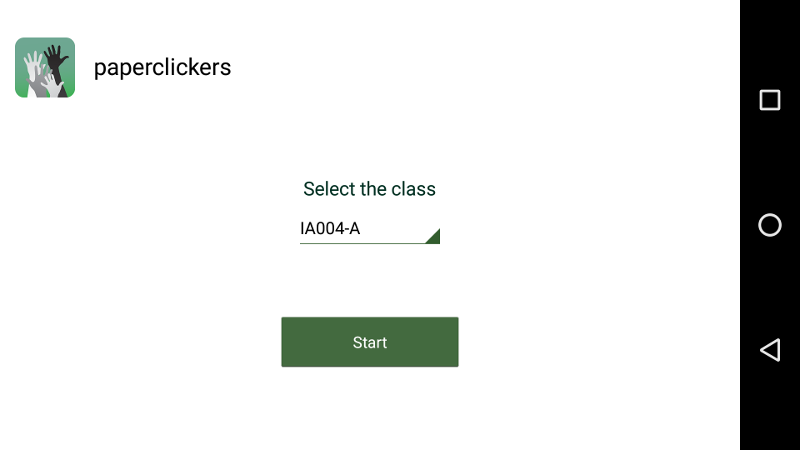}}
            \captionof{figure}{Initial screen -- original version}
            \label{fig:old_initialScreen}
        \end{minipage} &
        \begin{minipage}{0.5\textwidth}
            \centering
            \setlength{\fboxsep}{0pt}\fbox{\includegraphics[width=.9\linewidth]{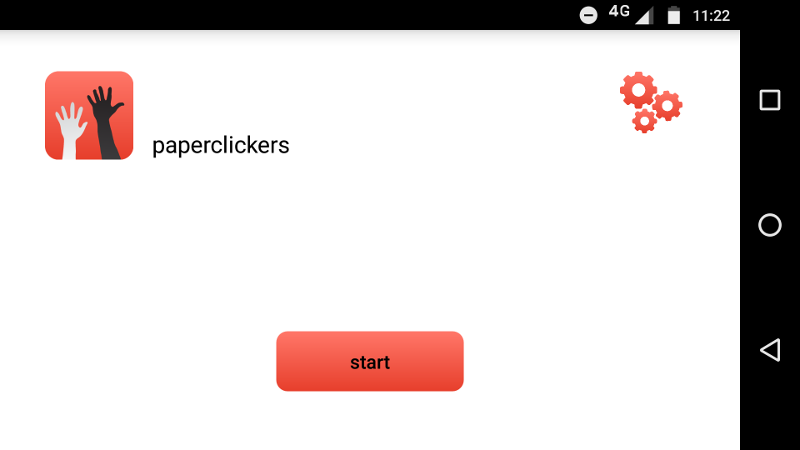}}
            \captionof{figure}{Initial screen -- released version}
            \label{fig:initialScreen}
        \end{minipage} \\
        \begin{minipage}{0.5\textwidth}
            \centering
            \setlength{\fboxsep}{0pt}\fbox{\includegraphics[width=.9\textwidth]{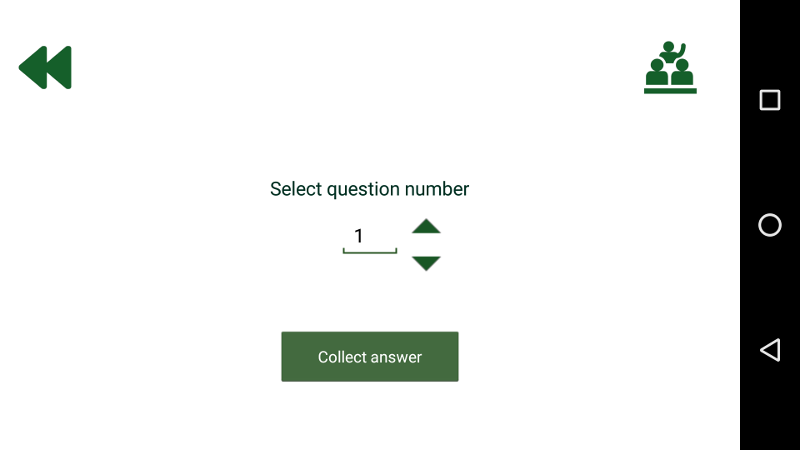}}
            \captionof{figure}{Question selection screen -- original version; roll call feature accessible in upper-right icon}
            \label{fig:old_questionSelection}
        \end{minipage} &
        \begin{minipage}{0.5\textwidth}
            \centering
            Screen removed on released version
        \end{minipage} \\
        \begin{minipage}{0.5\textwidth}
            \centering
            \setlength{\fboxsep}{0pt}\fbox{\includegraphics[width=.9\textwidth]{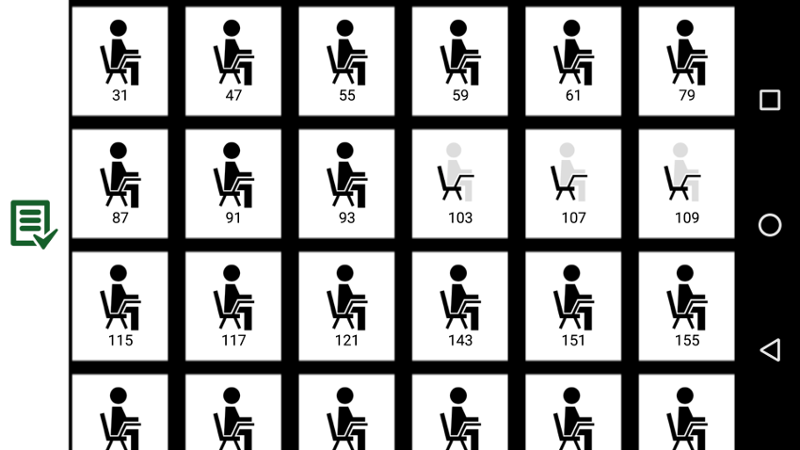}}
            \captionof{figure}{Roll call result screen -- original version}
            \label{fig:old_rollCallResult}
        \end{minipage} &
        \begin{minipage}{0.5\textwidth}
            \centering
            Screen removed on released version
        \end{minipage} \\
        \begin{minipage}{0.5\textwidth}
            \centering
            \setlength{\fboxsep}{0pt}\fbox{\includegraphics[width=.9\textwidth]{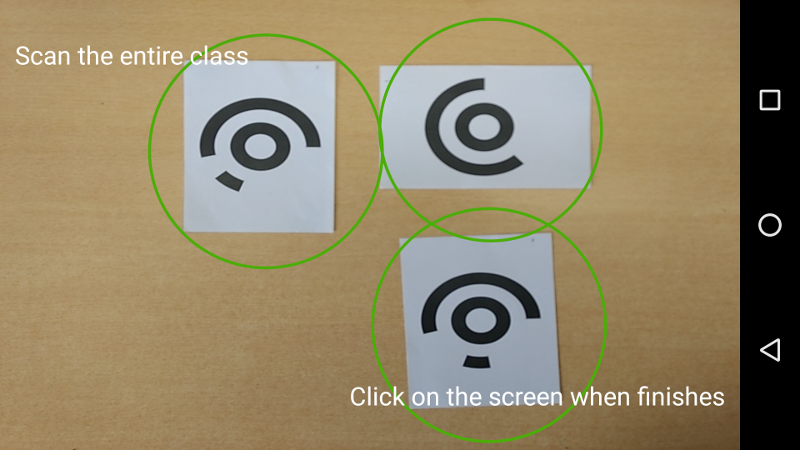}}
            \captionof{figure}{Scan screen -- original version}
            \label{fig:old_answersScan}
        \end{minipage} &
        \begin{minipage}{0.5\textwidth}
            \centering
            \setlength{\fboxsep}{0pt}\fbox{\includegraphics[width=.9\linewidth]{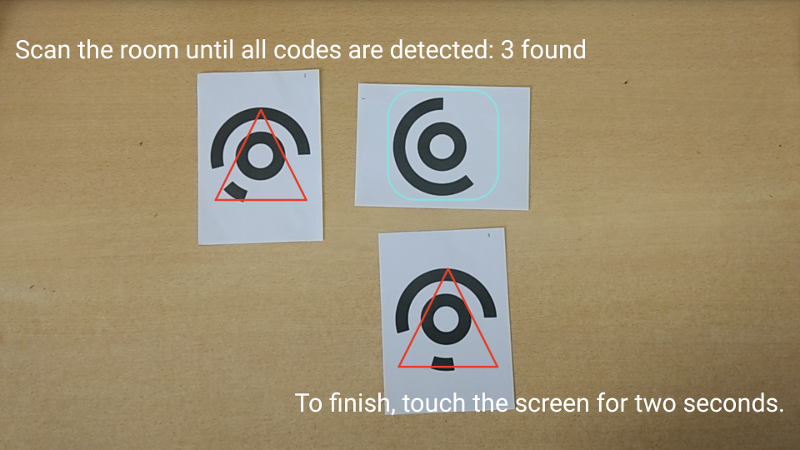}}
            \captionof{figure}{Scan screen -- released version}
            \label{fig:answersScan}
        \end{minipage} \\
    \end{tabular}
\end{table*}

\begin{table*}[p]
    \begin{tabular}[t]{m{0.5\textwidth}m{0.5\textwidth}}
        \begin{minipage}{0.5\textwidth}
            \centering
            \setlength{\fboxsep}{0pt}\fbox{\includegraphics[width=.9\textwidth]{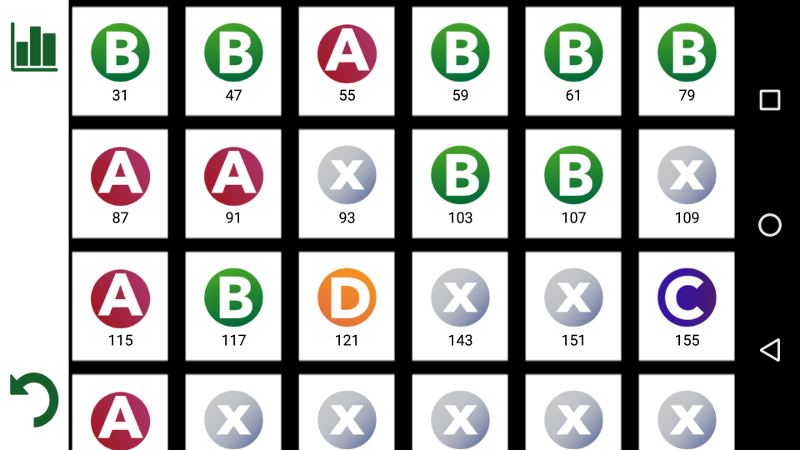}}
            \captionof{figure}{Detailed answers screen -- original version}
            \label{fig:old_detailedAnswers}
        \end{minipage} &
        \begin{minipage}{0.5\textwidth}
            \centering
            \setlength{\fboxsep}{0pt}\fbox{\includegraphics[width=.9\linewidth]{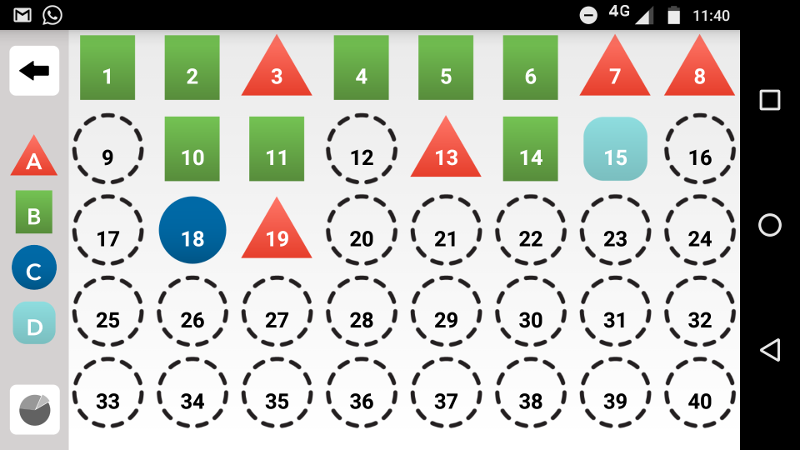}}
            \captionof{figure}{Detailed answers screen -- released version}
            \label{fig:detailedAnswers}
        \end{minipage} \\
        \begin{minipage}{0.5\textwidth}
            \centering
            \setlength{\fboxsep}{0pt}\fbox{\includegraphics[width=.9\textwidth]{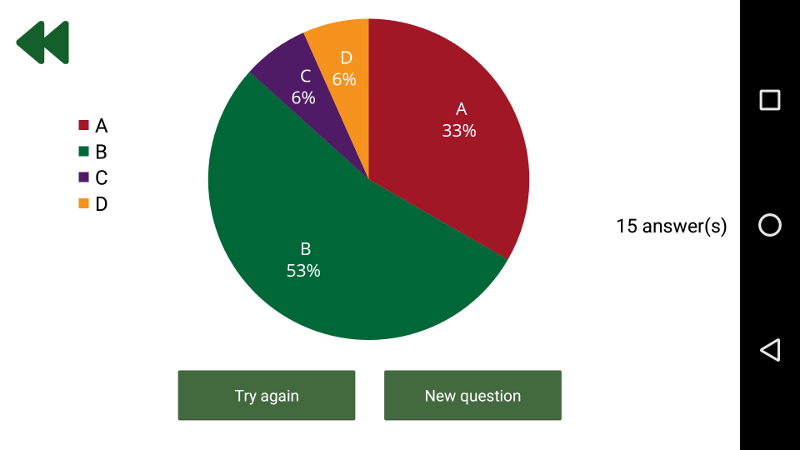}}
            \captionof{figure}{Chart screen -- original version}
            \label{fig:old_chart}
        \end{minipage} &
        \begin{minipage}{0.5\textwidth}
            \centering
            \setlength{\fboxsep}{0pt}\fbox{\includegraphics[width=.9\textwidth]{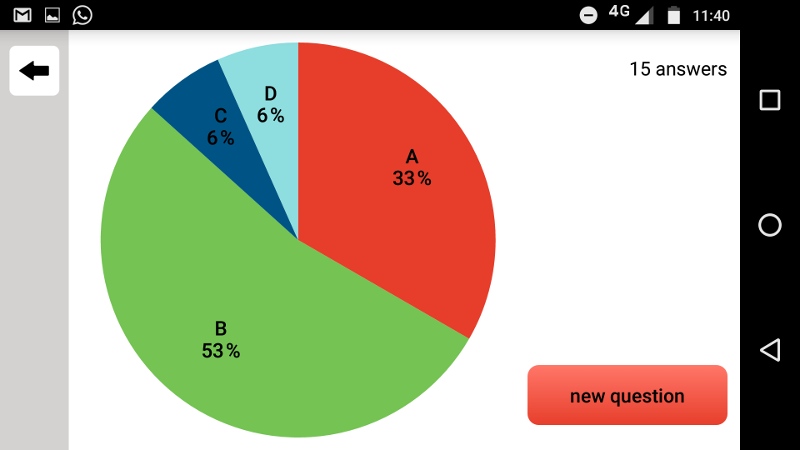}}
            \captionof{figure}{Chart screen -- released version}
            \label{fig:chart}
        \end{minipage} \\
        \begin{minipage}{0.5\textwidth}
            \centering
            Not included in original version
        \end{minipage} &
        \begin{minipage}{0.5\textwidth}
            \centering
            \setlength{\fboxsep}{0pt}\fbox{\includegraphics[width=.5\textwidth]{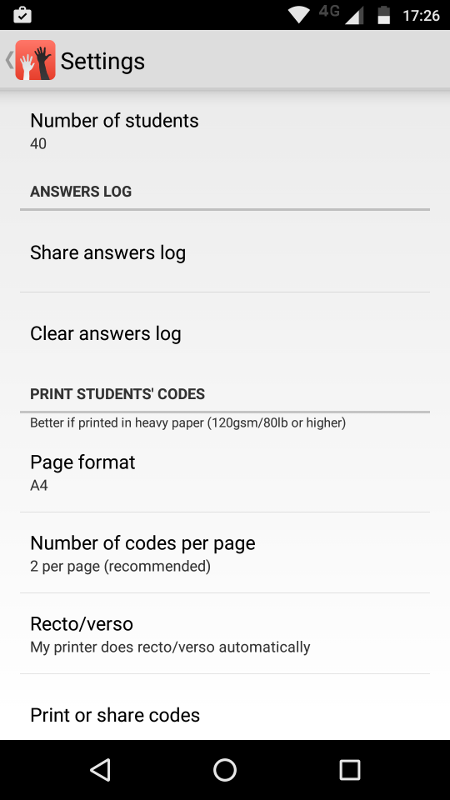}}
            \captionof{figure}{Settings screen -- released version}
            \label{fig:settings}
        \end{minipage} \\
    \end{tabular}
\end{table*}

\subsection{Second development cycle}
\label{sec:second_dev_cyvle}

The user experiments prompted us to perform an additional development cycle before the public release: not only there were important usability issues to work, but the speed and reliability of TopCodes detection could be improved. On that new phase we again relied on storyboards to redesign and guide the implementation changes. The second cycle concluded with the release of paperclickers on the Google Play Store\footnote{https://play.google.com/store/apps/details?id=com.paperclickers}, and the publication of its source code under the GNU Public Licence v2\footnote{https://github.com/learningtitans/paperclickers}.

The usability tests taught us that many extra features were more confusing than helpful, thus we turned to the Minimum Viable Product approach~\cite{ries2011lean}, focusing on the essential: collecting and summarizing students' responses. Keeping poll data for posterior analysis is still possible, but no longer encumbers the main workflow, if the teacher is interested in those extra features, she has to use the settings screen.


We removed the second screen entirely, along with its two features, the preliminary definitions (class and question) and roll call feature initialization. Although the users found those features interesting, they had trouble understanding and accessing them. 
We kept the main workflow as simple as possible for basic usage, and added a separate settings screen for advanced users, with options to add tags to the questions and to share the answers log, allowing those expert teachers to analyze individual students' answers. Figures~\ref{fig:old_initialScreen} until~\ref{fig:settings} contrast the main screens from the prototype \textit{versus} from the released versions.

\begin{figure}[ht]
    \centering
    \begin{minipage}{\linewidth}
        \centering
        \setlength{\fboxsep}{0pt}\fbox{\includegraphics[width=.9\textwidth]{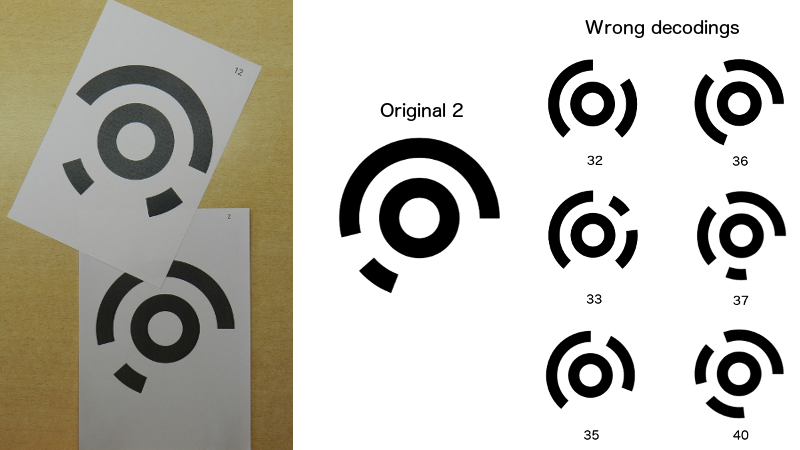}}
        \captionof{figure}{Decoding error due partial occlusion; on the right spurious detected codes for TopCode 2 partially occluded; in 123 scan cycles, TopCode 32 appeared in 0,82\% of the scan cycles, 33 in 0,82\%, 35 in 17,89\%, 36 in 8,94\%, 37 in 3,25\% and 40 in 8,13\% }
        \label{fig:decodingError}
    \end{minipage}
\end{figure}

We also worked to improve some issues found on the TopCodes detection and decoding. TopCodes original use case is the tracking of a small number of fiducial points to create virtual objects for virtual/augmented reality. They were not optimized for our context of tracking dozens of codes in very uncontrolled and dynamic environments, keeping careful track of each code orientation. In the second version, we found and corrected the problems explained below.

\textbf{Errors in detection and decoding:} In TopCodes' reference implementation, partially occluded codes are often decoded incorrectly (Figure~\ref{fig:decodingError}). In the original contexts, this creates a short transient detection that has no consequence other than quickly flashing an object out of place, but in paperclickers, errors of detection or decoding may set the wrong answer for a student. Worse: partial occlusions are very common as the teacher scans a classroom full of students. To overcome that issue, we added a time-consistency constraint to the detection: only codes which are detected across several contiguous video frames are considered valid; the threshold of frames is set dynamically according to the duration of the take.

\textbf{Too many code candidates:} TopCodes' reference implementation binarizes the image and then scans it horizontally, looking for the right transitions between black and white pixels, to find potential candidates for code locations. However, when the background has many vertical lines (e.g., curtains, blinds...), that results in a huge number of candidates, slowing the detection.  To reduce that sensitiveness to the background, we added a vertical scan step, and kept only the candidate points found both by the horizontal and the vertical scans.


\textbf{Sensitivity to hairline code defects:} Although TopCodes are extremely robust to affine transformations (scale, rotations, moderate camera baseline changes, etc.), we found them very sensitive to hairline defects, i.e., situations where a single row or column of the code becomes entirely white or black after binarization. We found those defects would be very common if the codes were printed in less-than-perfect printers, or if the students ignored the admonition to not fold the cards. After considering several complex solutions, we attempted using morphological operations to seal those small gaps. We tested many alternatives, but a binary closing followed by a binary opening using a 3 by 3 pixels square element offered the best compromise between eliminating defects and preserving details. However, further tests showed that the best solution --- both in terms of precision and speed --- was simply to instruct the user to not film from too close!

We also improved the overall detection speed --- one of the user complains about the prototype --- fine-tuning the grayscale conversion and using the Android's native image processing multi-core CPUs and GPUs usage\footnote{Android Renderscript computation engine framework - https://developer.android.com/guide/topics/renderscript}. After those changes, the TopCodes detection and decoding functionality reached the scan cycle performance of about 2 frames per second, including all the changes described above, running on 2017 mid-tier Android devices\footnote{1.5 GHz Cortex-A53 CPU, 1920x1080 pixels image}.

\section{Lessons learned} \label{sec:lessons}

The major challenge throughout development was ensuring a robust detection and decoding of the students' cards. Detecting and decoding a large number of cards in the uncontrolled environment of a classroom, while targeting low-cost computational device proved technically very challenging. Although TopCodes are very robust to distortions and noise, we had to create several additional adaptations to transpose them from their original application context (augmented reality) to ours (CRSs).

On the usability tests, the recording of the users interaction with the app --- including their ``think-aloud'' comments and  recommendations --- was the strategy that provided the most actionable information. The unstructured interviews were also interesting, but, to our surprise we found the the structured, formal survey the least useful of the instruments --- it only provided enough information to reinforce trends we had already understood --- with more confidence --- in the recordings and interviews. We believe that a survey has to be exceptionally well-designed to provide actionable information, while interviews and recording can be useful even for developers without a huge background in Human-Computer Interaction. In future projects, we will attempt to apply heuristic evaluation~\cite{nielsen1990heuristic} experiments with real users; we believe that cost-effective technique would have anticipated some of the problems found in our user trials.

Relying on storyboards for design and documentation worked very well for a small team,  designing a small-sized (less than 10-screen workflow) user-interaction driven application. Our team comprised 5 people, partially changing throughout the project --- a scenario not uncommon on academic research. We employed storyboards to elicit and document the requirements, to sketch the interaction elements, to design the navigation and dynamics of the application, etc. We also used them to image usage scenarios, which were also a crucial to design the usability tests.

\section{Conclusion} \label{sec:conclusion}

We proposed an effective working process to create a low cost classroom response system, based on recognizing fiducial markers as response devices. In order to release the solution, we have solved several technical issues due to reliability of fiducial markers detection in our application context, and usability challenges. The solution is available in both binary and source-code, as we hope it will be useful for other research teams interested in the subject.

Making available a low-cost educational CRSs is only part of what is needed to bring active learning to disfavored communities. It is critical to address teachers' and instructors' concerns related to adopting not only the tecnology, but mainly new teaching methods, especially when that represents leaving behind the safety, predictability and control of a lecture classroom setup~\cite{beatty2005transforming}. Our current work, thus, focus on creating pedagogical material for paperclickers, considering its usage by Brazilian teachers of Mathematics and Physics, in real Peer Instruction scenarios. We will also evaluate the user experience on the released paperclickers solution, including a diverse audience.



%



\ifCLASSOPTIONcompsoc
  \section*{Acknowledgments}
\else
  \section*{Acknowledgment}
\fi
We acknowledge the contributions of  Wilson Prata (funded by PIBIC/CNPq), and Vinícius Ribeiro (funded by PIBITI/CNPq) who helped conducting the user tests, and of Lucas Tejada who implemented the first idea of augmented reality in the first prototype.

Jomara Bindá was funded by FAPEAM project 003.2013,  Renato Lopes is funded by CNPq/PQ, and Eduardo Valle is funded by CNPq/PQ 311486/2014-2. The research was developed at RECOD Lab and at LCA Lab/UNICAMP which are funded by FAPESP, CAPES, and CNPq. Motorola Mobility Brazil, and Samsung Eletrônica da Amazônia kindly provided devices used to test the app. 

\ifCLASSOPTIONcaptionsoff
  \newpage
\fi



\bibliographystyle{IEEEtran}
\bibliography{IEEEabrv,references}
%

%



\begin{IEEEbiographynophoto}{Eduardo Oliveira}
received Bachelor degree in Computer Science and Social Sciences (political science, anthropology and sociology) from the University of Campinas (UNICAMP); his research interest is how computers and technology can compose educational technologies with proven social impact. He has been a software developer on the telecom industry for more than 20 years.
\end{IEEEbiographynophoto}

\begin{IEEEbiographynophoto}{Jomara Bindá}
is a motivated Ph.D. candidate researcher with a background in computer engineering. Her research primarily concerns to serve the community and promote social interactions using technology. With a particular focus on community health applications and information design, her work explores diverse HCI activities and service projects throughout the wider community. Jomara relishes academic research, and HCI in particular, as a field where a thorough understanding how people use technology to share and collaborate, to learn, to manage their lives, and even to be better citizens.
\end{IEEEbiographynophoto}

\begin{IEEEbiographynophoto}{Renato Lopes}
holds a degree (1994) and Master's degree (1997) in Electrical Engineering from the University of Campinas - UNICAMP and a PhD in Electrical Engineering from the Georgia Institute of Technology (2003). He is currently Assistant Professor at the University of Campinas - UNICAMP, working also as journal reviewer for several publications: IEEE Transactions on Communications, IEEE Transactions on Information Theory, EURASIP Journal on Applied Signal Processing, IEEE Transactions on Vehicular Technology, IEEE Journal on Selected Areas in Communications, Electronics Letters, IEEE Transactions on Signal Processing, IEEE Signal Processing Letters, IEEE Transactions on Wireless Communications, Reviewer IEEE Communications Letters. He is also member of the Editorial Board of the Journal of Communication and Information Systems (Online). He has experience in the Electrical Engineering Area, with emphasis on Telecommunications, acting mainly on the following topics: Minimization of Probability of Error, Iterative Techniques for Telecommunications, Maximum Likelihood, Blind Equalization.
\end{IEEEbiographynophoto}

\begin{IEEEbiographynophoto}{Eduardo Valle}
is a professor at the School of Electrical and Computing Engineering (FEEC) at the State University of Campinas (UNICAMP) since 2010. His research interest include Multimedia Information Retrieval, Content-based Information Retrieval, Large-Scale Machine Learning, Computer Vision. He is particularly interested in applications of Machine Learning for Education and Health.
\end{IEEEbiographynophoto}




\vfill


\end{document}